\documentclass[fleqn,twoside]{article}
\usepackage{espcrc2}

\newcommand{\AmS}{{\protect\the\textfont2
  A\kern-.1667em\lower.5ex\hbox{M}\kern-.125emS}}

\def \b{\mathcal{B}}
\def \beq{\begin{equation}}
\def \bnl{Brookhaven National Laboratory}
\def \eeq{\end{equation}}
\def \ket#1{| #1 \rangle}
\def \pb{presented by~}
\def \s{\sqrt{2}}
\def \sst{\sin^2 \theta}
\def \st{\sqrt{3}}
\def \sx{\sqrt{6}}
\def \tc{this conference}

\def \arnps#1#2#3{Ann.~Rev.~Nucl.~Part.~Sci. #1 (#3) #2}

\def \ba88{Particles and Fields 3 (Proceedings of the 1988 Banff Summer
Institute on Particles and Fields), edited by A. N. Kamal and F. C. Khanna
(World Scientific, Singapore, 1989)}

\def \be87{Proceedings of the Workshop on High Sensitivity Beauty
Physics at Fermilab, Fermilab, Nov. 11--14, 1987, edited by A. J. Slaughter,
N. Lockyer, and M. Schmidt (Fermilab, Batavia, IL, 1988)}

\def \cn{Collaboration}

\def \cp89{{\it CP Violation,} edited by C. Jarlskog (World Scientific,
Singapore, 1989)}

\def \epja#1#2#3{Eur.~Phys.~J. A #1 (#3) #2}

\def \hb87{Proceeding of the 1987 International Symposium on Lepton and
Photon Interactions at High Energies, Hamburg, 1987, ed. by W. Bartel
and R. R\"uckl (Nucl.~Phys.~B, Proc. Suppl., vol. 3) (North-Holland,
Amsterdam, 1988)}

\def \ite{{\it et al.}}

\def \jpg#1#2#3{J. Phys.~G #1 (#3) #2}

\def \lkl87{Selected Topics in Electroweak Interactions (Proceedings of 
the Second Lake Louise Institute on New Frontiers in Particle Physics, 15--21
February, 1987), edited by J. M. Cameron \ite~(World Scientific, Singapore,
1987)}

\def \np#1#2#3{Nucl.~Phys. #1 (#3) #2}

\def \PDG{Particle Data Group, K. Hagiwara \ite, \prd{66}{010001}{2002}.}
 
\def \pl#1#2#3{Phys.~Lett. #1 (#3) #2}

\def \plb#1#2#3{Phys.~Lett. B #1 (#3) #2}

\def \prd#1#2#3{Phys.~Rev. D #1 (#3) #2}
\def \prl#1#2#3{Phys.~Rev.~Lett. #1 (#3) #2}

\def \si90{25th International Conference on High Energy Physics, Singapore,
Aug. 2-8, 1990, Proceedings edited by K. K. Phua and Y. Yamaguchi (World
Scientific, Teaneck, N. J., 1991)}

\def \yaf#1#2#3#4{Yad.~Fiz. #1 (#3) #2 [Sov.~J.~Nucl.~Phys.,~#1 (#3) #4]}

\title{Conference Summary}

\author{Jonathan L. Rosner
\address{Enrico Fermi Institute and Department of Physics, University of
 Chicago \\ 
 5640 S. Ellis Avenue, Chicago, IL 60637}
        \thanks{Presented at 5th International Conference on Hyperons,
                Charm, and Beauty Hadrons, Vancouver, BC, Canada,
                25--29 July 2002.  EFI 02-57, hep-ph/0208243.}}

\begin{document}

\begin{abstract}
A summary is given of the 5th International Conference on Hyperons, Charm and
Beauty Hadrons held in Vancouver, Canada, June 25th to 29th, 2002.  This series
of conferences began in 1995 in Strasbourg, France, in large part through the
efforts of A. Fridman, to whose memory this talk is dedicated.  Topics reviewed
include kaon and hyperon physics, charm and beauty production and decays, heavy
baryons, the physics of the Cabibbo-Kobayashi-Maskawa matrix and CP violation,
and precision electroweak analyses.  An attempt is made to combine a review of
the high points of the conference with a more general overview of the field and
its prospects.
\vspace{1pc}
\end{abstract}

\maketitle


\section{INTRODUCTION AND DEDICATION}

This series of conferences was begun in Strasbourg in 1995, largely through the
efforts of A. Fridman, or ``Fredy,'' as he was known, a remarkable individual
whom I came to know during one of his visits to Tel Aviv University in 1968
and later when we were both at CERN.  My family and I have fond memories of
his generous hospitality in Strasbourg in 1973, when he treated us at a
restaurant in the nearby French countryside to one of the best meals I have
ever had.  Fredy cared deeply about physics and had particular tastes
(such as heavy baryons) which ensured that topics which were not always
fashionable received the attention they deserved.  We miss him greatly.

The start of my visit to Canada speaks well for the visibility of particle
physics here.  At the border the guard asked my wife and me for photo
identification and for the purpose of our visit.  When I named the conference,
he asked:  ``What's your favorite subatomic particle?  Do you think dark
matter will be found?  Will the Universe keep expanding or collapse back to a
point?  Personally, I like the cyclic idea.'' I asked him if he wanted to see
our proof of citizenship; he answered:  ``Nope.  Have a nice day!''  We later
learned that the press coverage of particle physics in Vancouver (and  the rest
of Canada) has been extensive, and that the guards are quite well-informed.

This Conference has been an enjoyable mixture of topics which, while not
all answering the border guard's deep questions, shed light on many fundamental
issues, such as the pattern of quark masses and mixings, the origin of the CP
violation in the kaon and $B$ meson systems, and possibilities for physics
beyond the Standard Model.  I first discuss the
unitarity of the Cabibbo-Kobayashi-Maskawa (CKM) matrix as probed via the
first two quark families (Sec.\ \ref{sec:un}), and then kaons (Sec.\
\ref{sec:K}) and hyperons (Sec.\ \ref{sec:hyp}).  A trio of sections is
devoted to spectroscopy:  quarkonium (Sec.\ \ref{sec:qk}), particles with
charm (Sec.\ \ref{sec:c}) and particles with beauty (Sec.\ \ref{sec:b}).
In the last two of these I also discuss weak decays and what they teach us
about the strong and weak interactions and the CKM matrix.
A separate section (\ref{sec:hqp}) is devoted to heavy quark production.
I then treat the electroweak sector in Sec.\ \ref{sec:ew}, including the
neutral-current couplings of heavy quarks and the search for the Higgs boson.
Physics beyond the Standard Model (mainly supersymmetry) is discussed in
Sec.\ \ref{sec:bsm}, while Sec.\ \ref{sec:pros} concludes.

\section{CKM AND THE FIRST TWO FAMILIES}
\label{sec:un}

Is the CKM matrix unitary \cite{Manohar}?  Superallowed transitions in nuclei
\cite{GKR} yield $|V_{ud}| \simeq 0.9740(5)$, while the lifetime and $g_A/g_V$
of the neutron yield a slightly smaller value $\sim 0.973$ \cite{nVud}.  The
value of $|V_{us}|$ quoted for many years \cite{GKR}, based on $K_{e3}$ decays,
has been $\sim 0.220 \pm 0.002$, yielding $\sum_i|V_{ui}|^2 = 0.996 \pm 0.002$,
a $2 \sigma$ discrepancy.  At \tc~we heard of a new analysis of hyperon decays
\cite{Swallow} which gives $|V_{us}| = 0.2250 \pm 0.0027$, while a new
$K_{e3}$ experiment \cite{E865} is likely to give about a 3\% increase
in the old value of $|V_{us}|$ modulo radiative corrections \cite{Bytev}.

The value of $|V_{cd}|$ is roughly 0.22, in accord with unitarity expectations
\cite{GKR}.  At \tc~we heard of a new value of $|V_{cs}|$ extracted from
$W$ decays \cite{Ealet}.  Using precision measurements and theoretical
estimates of the $W$ production cross section, the LEP II collaborations have
measured $\b (W \to {\rm hadrons}) = (67.92 \pm 0.27)\%$ (67.5\% in
the Standard Model [SM]) and $\b (W \to l \nu) = (10.69 \pm 0.09)\%$
(10.8\% in the SM), where the SM predictions are based on $\alpha_s(M_W) =
0.121 \pm 0.002$.  This constrains $\sum|V_{ij}|^2$, where $i=u,c;~j=d,s,b$.
Subtracting known values, one finds $|V_{cs}| = 0.996 \pm 0.013$, whereas
unitarity would predict $\simeq 0.975$.  The agreement is better than
$2 \sigma$.

\section{KAONS}
\label{sec:K}

The KLOE detector at the DAFNE electron-positron collider at Frascati has
been studying $e^+ e^- \to \phi \to  \ldots$, where in addition
to $K \bar K$ the final states include $f_0 \gamma$ and $\eta' \gamma$.
The newly measured branching ratio $\b (\phi \to \eta' \gamma) = (6.8 \pm
0.6 \pm 0.5) \times 10^{-5}$ probes the strange quark content of the $\eta'$
\cite{JRetap}.  Writing $\ket{\eta'} = X \ket{u \bar u + d \bar d}/\s +
Y \ket{s \bar s} + Z \ket{{\rm glue}}$, one finds that $X^2 + Y^2 = 0.95
^{+0.11}_{-0.07}$, limiting the amount of glue in the $\eta'$ wave function
\cite{Branchini}.

Two events of the rare decay $K^+ \to \pi^+ \nu \bar \nu$ have now been seen
\cite{Numao,E787}, corresponding to $\b = (1.57^{+1.75}_{-0.82}) \times
10^{-10}$.  This is to be compared with the SM prediction of $\b = (0.82 \pm
0.32) \times 10^{-10}$; the amplitude probes a combination proportional to
$|1.4 - \rho - i \eta|$ of the Wolfenstein \cite{WP} parameters $\rho$ and
$\eta$.  The goals of BNL experiment E949 \cite{Numao} and the Fermilab CKM
experiment \cite{Nguyen} are to record 10 and 100 events of this process,
respectively, if the SM prediction is correct.

Events have been seen in $K_L^0 \to \pi^0 e^+ e^-$ and in $K_L^0 \to \pi^0
\mu^+ \mu^-$ at a level consistent with background \cite{Monnier}. Further
study of radiative $K_L^0$ decays (e.g., to $\gamma \gamma e^+ e^-$) and a
search for $K_S \to \pi^0 e^+ e^-$ will provide useful information.  The
present limit is $\b(K_S \to \pi^0 e^+ e^-) < 1.4 \times 10^{-7}$ \cite{Sozzi}.
The CP-conserving amplitude for this process is fed by rescattering from
$K_L \to \pi^0 \gamma \gamma$, for which NA48 has presented new results
\cite{Sozzi,pigg}.  The amplitude for $K_L^0 \to \pi^0 \nu \bar \nu$ is
proportional to the CP-violating parameter $\eta$.  The SM prediction $\b =
(3.1 \pm 1.3) \times 10^{-11}$ will be approached stepwise, with experiments at
KEK (E391), the Japan Hadron Facility (JHF) and BNL (KOPIO) \cite{Numao}.

The parameter $\epsilon'/\epsilon$ in $K_{S,L} \to \pi \pi$ is still evolving.
Fermilab E832 presented its value based on the full 1996-7 data set,
$\epsilon'/\epsilon = (20.7 \pm 2.8) \times 10^{-4}$ \cite{Monnier,%
Alavi:2002}.  A new value from CERN NA48, $\epsilon'/\epsilon = (14.8 \pm 2.2)
\times 10^{-4}$ \cite{Sozzi,Batley}, when combined with the Fermilab result,
yields a world average $\epsilon'/\epsilon = (17.0 \pm 2.9) \times 10^{-4}$,
where I have increased the error from $\pm 1.7$ by $S \equiv (\chi^2)^{1/2}$
\cite{PDG}.  The Fermilab experiment's results of its 1999 run should reduce
the error further.

The amplitudes contributing to $K_{S,L} \to \pi \pi$ consist of a penguin
(with a weak phase, leading only to an $I=0$ final state) and a tree (with
no weak phase, leading to both $I=0$ and $I=2$ final states).  The relative
phase of the $I=2$ and $I=0$ amplitudes leads to $\epsilon' \ne 0$.
Despite the difficulty of estimating the relative strength of penguin and
tree contributions, the Fermilab and CERN measurements definitely establish
the presence of direct CP violation.

Other rare $K$ decay process include $K_L \to e^+ e^- \gamma$ and $K_L \to
\mu^+ \mu^- \gamma$, whose dependence on $m(l^+l^-)$ probes the form factor
in $K_L \to \gamma^* \gamma$, useful for estimating the contribution of
the long-distance dispersive contribution (and hence also the short-distance
contribution) to $K_L \to \mu^+ \mu^-$ \cite{Monnier}.  It now appears that
the $e^+ e^- \gamma$ and $\mu^+ \mu^- \gamma$ results give the same
form-factor parameters, which was not always the case.

\section{HYPERONS}
\label{sec:hyp}

The neutral hyperons produced in neutral kaon beams have been studied
at Fermilab \cite{Swallow,Monnier} and CERN \cite{Munday}.  The decay
$\Xi^0 \to \Sigma^+ e^- \bar \nu_e$ is related to $n \to p e^- \bar \nu_e$
by the U-spin transformation $d \leftrightarrow s$; differences between the
two probe SU(3) breaking.  The Fermilab group \cite{Monnier} has recently
studied a large sample of events for this process.  The CERN group
\cite{Munday} has measured the $\Xi^0$ mass much more precisely than
previously, allowing a test of the Coleman-Glashow relation $m(n) - m(p) +
m(\Xi^-) - m(\Xi^0) = m(\Sigma^-) - m(\Sigma^+)$.  This relation should be good
to $0.1$ MeV \cite{Manohar,Jenkins}, as it seems to be.

Radiative hyperon decays \cite{Zenc} have posed a long-standing puzzle.
Up to now we have not had a consistent description of all rates and
polarization asymmetries for the processes $\Sigma^+ \to p \gamma$,
$\Lambda \to n \gamma$, $\Xi^0 \to (\Lambda,\Sigma^0) \gamma$, $\Xi^- \to
\Sigma^- \gamma$, and $\Omega^- \to \Xi^- \gamma$.  Early work \cite{GW}
showed that an elementary $s \to d \gamma$ transition was not enough, while
a quark-model treatment \cite{GLOPR} appeared to disagree with data.  Recent
CERN NA48 data on the asymmetry parameter in $\Xi^0 \to \Lambda \gamma$
seems to have resolved the question in favor of the general scheme of
predictions of the quark model \cite{GLOPR} and in favor of Hara's
Theorem \cite{Hara}, which says that polarization asymmetries in these
radiative decays should vanish in the limit of exact SU(3).

The HyperCP Experiment at Fermilab has provided new information on charged
hyperons.  The decay $\Omega^- \to \Xi^- \pi^+ \pi^-$ has been oberved with
$\b = (3.6 \pm 0.3) \times 10^{-4}$ \cite{Solomey}.  The mechanism is still
unclear; should one see a $\Xi^*(1530)~(3/2^+)$ intermediate state?  The
T-violating asymmetry in $\Xi^- \to \Lambda \pi^- \to p \pi^- \pi^-$ has been
bounded \cite{Zyla}:  $A_{\Xi \Lambda} = (- 7 \pm 12 \pm 6.2) \times 10^{-4}$,
with an error of $2 \times 10^{-4}$ (still above SM expectations) expected
for the full sample.

\section{QUARKONIUM}
\label{sec:qk}

The Belle Collaboration
has discovered the radial excitation of the $^1S_0$ quarkonium ground state,
the $\eta_c'(2S)$ with a mass of 3654 MeV \cite{etacp} through the
decay $B \to \eta_c'(2S) K \to K_S K^\pm \pi^\mp K$.  The $2S$ hyperfine
splitting $\psi' - \eta_c' = 32$ MeV is considerably less than the $1S$
splitting $J/\psi - \eta_c = 118$ MeV and less than one would estimate on
the basis of the $J/\psi$ and $\psi'$ leptonic widths, suggesting the
possibility that coupled-channel effects and/or mixing with the $\psi''(3770)$
are pushing down the $\psi'$ mass \cite{MR82,psip,ELQ}.

The search
for singlet P-wave $c \bar c$ and $b \bar b$ states has not yet produced a
firm candidate.  The Fermilab E835 Collaboration \cite{Rumerio} still is mute
regarding their previous (E760) claim for a $h_c \equiv c \bar c(^1P_1)$
state at 3526 MeV.  Suzuki \cite{Suz02} has suggested looking for $B \to
h_c K$, with $h_c \to \eta_c \gamma$.  As for the $h_b = b \bar b(^1P_1)$,
which should decay largely to $\gamma \eta_b$, it should be produced via
$\Upsilon(3S) \to \pi^0 h_b$ \cite{Voloshin} or $\Upsilon(3S) \to \pi^+ \pi^-
h_b$ \cite{KY}.  For other suggestions for P-wave singlet observation, and a
summary of mass predictions, see \cite{GRP}.

The D-wave quarkonium levels \cite{KR,Chao,GRD} include the observed $c \bar c
(^3D_1)$ state $\psi''(3770)$ and a candidate for a $b \bar b(^3D_2)$ state
$\Upsilon(10162)$ reported since this conference by the CLEO \cn~at ICHEP2002
in Amsterdam \cite{CLEOD}. Chao \cite{Chao} has suggested that the $c \bar c
(1D_2)$ state may be narrow and easily observed.

In light of new data on quarkonium decays (e.g., \cite{Rumerio,Zhu}), new
approaches based on non-relativistic QCD (NRQCD) \cite{Brambilla,Pineda,%
Sanchis,Vairo}, and realization of the importance of color octet contributions
to quarkonium wave functions, previous $\alpha_s$ determinations based on heavy quarkonium decays to light hadrons (e.g., \cite{KMRR}) should be updated.
The BES group \cite{Zhu} has presented impressive evidence that
$f_J(1710)$ produced in $J/\psi \to f_J \gamma$ has $J=0$; it is a leading
candidate for the lightest glueball.

\section{PARTICLES WITH CHARM}
\label{sec:c}

Candidates for doubly-charmed baryons presented by the Fermilab SELEX
\cn~\cite{Cooper,ccq} are summarized in Table 1.  These entail several
mysteries.

\begin{table}
\caption{SELEX Candidates for doubly-charmed baryons.}
\begin{tabular}{c c c c} \hline \hline
 State  & Mass  & Sig. & Mode \\
        & (MeV) &      &      \\ \hline
$ccd^+$ &    3519    & $6.3 \sigma$ & $\Lambda_c^+ K^- \pi^+$ \\
$ccu^{++}$ & 3460    & $4.8 \sigma$ & $\Lambda_c K^- \pi^+ \pi^+$ \\
$ccu^{*++}$ & 3783   & $4.0 \sigma$ & $\Lambda_c K^- \pi^+ \pi^+$ \\ \hline
\hline
\end{tabular}
\end{table}

\begin{itemize}

\item The FOCUS \cn~\cite{Ratti} does not see a signal, though they detect
many more $\Lambda_c$'s.

\item The isospin splitting between the first two states is enormous.
One usually expects isospin splittings to be at most a few MeV.

\item The production rate of the doubly-charmed baryons is so large that
fully half of all observed $\Lambda_c$'s must come from their decays.

\item The hyperfine splitting between the last two states (presumably
$J=1/2$ and $J=3/2$ candidates) is larger than one might estimate by
elementary means (see, e.g., \cite{DGG}). 

\end{itemize}

Impressive results on charmed particle photoproduction have been
presented by the FOCUS (E831) Collaboration at the Fermilab Tevatron.

\begin{table*}
\caption{Charmed particle lifetimes as of June 2002.}
\begin{tabular}{c c c c c c c c} \hline \hline
$\tau$(fs) & $D^+$ & $D^0$ & $D_s^+$ & $\Xi_c^+$ & $\Lambda_c^+$ & $\Xi_c^0$ &
 $\Omega_c^0$ \\ \hline
FOCUS & 1039.4 & 409.6 & 506 & 439 & 204.6 & 118 & 79 \\
      & $\pm 4.3 \pm 12$ & $\pm 1.1 \pm 1.5$ & $\pm 8$ & $\pm 22 \pm 9$ &
 $\pm 3.4 \pm 2.5$ & $^{+14}_{-12} \pm 5$ & $\pm 12$ \\ \hline
PDG   &  1051  & 411.7 & 490 & 442 &  200  &  98 & 64 \\
2002  & $\pm 13$  & $\pm 2.7$ & $\pm 9$ & $\pm 26$ &  $\pm 6$  & $^{+23}_{-15}$
 & $\pm 20$ \\ \hline \hline
\end{tabular}
\end{table*}

{\it Charmed particle semileptonic decays} \cite{Lopez} have yielded evidence
for an S-wave $K \pi$ contribution under the dominant $\bar K^*(890)$ vector
meson resonance in $D^+ \to K^- \pi^+ \mu^+ \nu_\mu$.  The branching ratio
$\b(D^+ \to \bar K^{*0} \mu^+ \nu_\mu) = (5.5 \pm 0.4)\%$ is larger than
the Particle Data Group \cite{PDG} value of $(4.8 \pm 0.4)\%$ but $1.6 \sigma$
below the CLEO \cite{CLEOsl} value of $(6.7 \pm 0.8)\%$.  It was always
puzzling why this branching ratio was not larger \cite{AR}.  The new FOCUS
measurement $\b(D_s \to \phi \mu^+ \nu_\mu)/\b(D_s \to \phi \pi^+) =
0.54 \pm 0.06$ can be helpful in calibrating absolute $D_s$ branching ratios
if supplemented by theoretical estimates of the $D_s \to \phi$ form factors.

{\it Charmed baryon decays} \cite{Ratti} have been observed by FOCUS in several
Cabibbo-suppressed $\Lambda_c^+$ modes, including $\Sigma^+ \bar K^{*0}$
(strangeness $S=-2$), $\Sigma^+ K^{*0}~(S=0)$, and $\Sigma^- K^+ \pi^+~(S=0)$.
Cabibbo-favored $S=-1$ modes such as $\Sigma^+ \pi^+ \pi^-$ and $\Sigma^+ K^+
K^-$ also have been observed; in the latter case the Dalitz plot shows a
$\phi$ band in $K^+ K^-$ and a $\Xi^*$ band in $\Sigma^+ K^-$.  As mentioned,
there are no signs of SELEX's $ccq$ baryons.

{\it Charmed particle lifetimes} \cite{Vaan} exhibit a hierarchy which is
qualitatively understood from the standpoint of heavy quark symmetry.  In
Table 2 we compare the FOCUS results with Particle Data Group (2002) averages.
The FOCUS $D_s$ and $\Omega_c$ values are preliminary.

The CLEO \cn~\cite{PedlarD} has presented results on $D^0 \to K_s
\pi^+ \pi^-$ in which the doubly-Cabibbo-suppressed decay $D^0 \to K^{*+}
\pi^-$ is seen via interference on the Dalitz plot.  Through a study of related
decay modes one can learn about final-state phase differences \cite{CRD} in
the same way as for the Cabibbo-favored modes \cite{JRcharm}.

Several presentations at this Conference dealt with charmed meson spectroscopy
and rare decays.  Predictions of the lowest $ccq$ state differed greatly:
3241 MeV \cite{Mat} or 3640--3690 MeV \cite{Nar}, where both values refer
to the spin-weighted $J=1/2$ and $J=3/2$ average.  The latter is
consistent with the SELEX \cite{Cooper} claim.  It would be helpful to have
predictions of hyperfine splittings to compare with SELEX's large value.
S. Fajfer \cite{Faj} pointed out that in $D \to V \gamma$ and
$D \to \gamma \gamma$, long-distance effects are likely to dominate, with
$\b(D^0 \to \gamma \gamma) \simeq 2^{\pm 1} \times 10^{-8}$ and $\b(D_s^+
\to \rho^+ \gamma) \simeq 2^{\pm 1} \times (4 \times 10^{-4})$.  Short-distance
effects could lead to a difference between $\b(D^0 \to \rho \gamma)$ and
$\b(D^0 \to \omega \gamma)$.

\section{PARTICLES WITH BEAUTY}
\label{sec:b}

\subsection{Semileptonic decays}

\subsubsection{Inclusive $b \to c$ transitions}

The leading-order expression for the semileptonic $b \to c$ decay width is
\beq
\Gamma_0 = \frac{G_F^2 m_b^5}{192 \pi^3} |V_{cb}|^2 f(m_c/m_b)~~~,
\eeq
where $f(x)$ is a known function equal to 1 for $x = 0$ and about 1/2 for $x =
m_c/m_b$.  This expression is clearly very sensitive to $m_b$, though less so
when $m_b - m_c$ is confined within its known limits (about 3.34 to 3.4 GeV).
As a result of work reported at this conference \cite{Bauer,Chistie,Jin,%
Urheim,%
Waller}, the $m_b$ dependence is being tamed; information on $b \to s \gamma$
is helpful.  Calculations of the semileptonic spectra agree with experiment
and can be used for baryons as well \cite{Kalman}.  One finds $|V_{cb}| \simeq
0.041$ with about a 5\% error.  There seems to be no missing charm problem
in $B$ decays \cite{Waller}.

\subsubsection{Exclusive $b \to c$ transitions}

The measurement of the spectrum in $B \to D^* l \nu$ and extrapolation to the
zero-recoil (maximum $m_{l \nu}$ mass) point now has been performed by a number
of groups, leading to a world average $\mathcal{F}(1)|V_{cb}| = (37.8 \pm
1.1) \times 10^{-3}$.  CLEO and ALEPH find values somewhat larger and smaller,
respectively, than this average, which also contains values from Belle,
DELPHI, and OPAL.  The most recent estimate from lattice gauge theory is
$\mathcal{F}(1) = 0.919^{+0.030}_{-0.035}$, so that this exclusive method
again gives $|V_{cb}| \simeq 0.041$ with about a 5\% error.

\subsubsection{Inclusive $b \to u$ transitions}

The decay $B \to X_u l \nu$ is plagued by background from $B \to X_c l \nu$,
which occurs with about 50 times the rate.  There are several ways to isolate
the desired signal \cite{Elias,Luke,Kow}.  The best is to make cuts in both
$q^2 = m^2_{l \nu}$ and $M_X$, but this requires neutrino reconstruction.  Next
best, in that order, are cuts in $q^2$, $M_X$, and the lepton energy $E_l$,
but this is exactly opposite to experimental feasibility!

At present CLEO \cite{Urheim} has used an inclusive method to determine
$|V_{ub}| \simeq 0.10 |V_{cb}|$ with about a 15\% error; reduction
to 10\% seems feasible.

\subsubsection{Exclusive $b \to u$ transitions}

The decays $B \to (\pi,\rho,\omega) l \nu$ can provide information on
$|V_{ub}|$ when form factors are specified (e.g., through lattice gauge theory
calculations).  A value based on Belle data was presented by Piilonen
\cite{Piil}.  The study of $B \to \pi l \nu$, in particular, is helpful in
estimating the contribution of the tree amplitude in $B \to \pi \pi$ if
factorization is assumed \cite{LRpipi}.

\subsection{$B \to D \pi$ isospin triangle}

A year ago Belle and CLEO reported observation of some color-suppressed
$D$ decays, including $B^0 \to \bar D^0 \pi^0$.  The rate for this process was
found to be sufficiently large that the triangle of complex amplitudes
for $B^0 \to \bar D^0 \pi^0$, $B^0 \to D^- \pi^+$, and $B^+ \to \bar D^0 \pi^+$
appeared to have non-zero area.  The branching ratios for the last two
processes were based on a sub-sample of the CLEO data.  Now CLEO has reported a
new analysis of the last two modes \cite{PedlarD,Ahmed}, which strengthens the
argument for a non-zero
final state phase difference between the $I = 1/2$ and $I = 3/2$ amplitudes.
Defining $\delta_I = {\rm Arg}(A_{3/2}/A_{1/2})$ (using the convention of
\cite{JRcharm}), and the new branching ratios $\b(B^0 \to D^- \pi^+) =
(26.8 \pm 2.9) \times 10^{-4}$, $\b(B^+ \to \bar D^0 \pi^+) = (49.7 \pm 3.8)
\times 10^{-4}$ as well as the Belle-CLEO average $\b(B^0 \to \bar D^0 \pi^0)
= (2.92 \pm 0.45) \times 10^{-4}$, one finds $\delta_I \simeq 30^\circ$,
$16^\circ \le \delta_I \le 33^\circ$ with 90\% confidence, or $\cos \delta_I <
1$ at $2.4 \sigma$.

Non-zero final-state phases in $B$ decays would be useful in observing direct
CP violation.  No such effects have been seen yet; the closest is in
$B \to \pi \pi$, where Belle sees a direct CP asymmetry but BaBar does not
(see below).

\subsection{Decays involving $\eta$ and $\eta'$}

A review of $B$ decays involving $\eta$ and $\eta'$ was presented by
B. Brau \cite{Brau}.  Several speakers \cite{Gard,Lip02,OD} were concerned 
with whether the large branching ratio $\b(B \to K \eta') \simeq 63
\times 10^{-6}$ represents a serious challenge to theory.  The standard penguin
amplitude may be represented by $P$, and an additional ``singlet penguin''
contribution, in which the $\eta$ or $\eta'$ couples to the process in a manner
violating the Okubo-Zweig-Iizuka (OZI) rule.  We represent
\beq
\label{eqn:wfs}
\eta \simeq (u \bar u + d \bar d - s \bar s)/\st~~,
\eeq
\beq
\eta' \simeq (u \bar u + d \bar d + 2 s \bar s)/\sx~~.
\eeq
Then (neglecting a small tree contribution to the charged $B$ decay)
\beq
A(B \to \eta K)  = 0 \cdot P + \frac{1}{\st} S~~,
\eeq
\beq
A(B \to \eta' K) = \frac{3}{\sx} P + \frac{4}{\st} S~~.
\eeq
The penguin contributions of nonstrange and strange quarks interfere
destructively in $\eta K$ and constructively in $\eta' K$ \cite{Lip87}.
They cancel completely for the particular mixing chosen in (\ref{eqn:wfs}),
which corresponds to octet-singlet mixing with an angle of $19^\circ$.

Calibrating the strength of the penguin amplitude with the process
$B^+ \to \pi^+ K^0$, whose branching ratio is $\b \simeq 17 \times 10^{-6}$,
one predicts the branching ratios shown in Table 3.  A small singlet ($S$)
contribution is needed to obtain the observed $\eta' K$ branching ratio.
This contribution is larger than most perturbative QCD estimates, but is
obtained satisfactorily by Beneke \ite~in their generalized factorization
approach \cite{Ben02}.

\begin{table}
\caption{$B$ branching ratios as function of $S/P$.}
\begin{tabular}{c c c} \hline \hline
$S/P$ & $\b(\eta' K)$ & $\b(\eta K)$ \\ \hline
   0  & $26 \times 10^{-6}$  &           0        \\
  0.4 & $60 \times 10^{-6}$ & $0.9 \times 10^{-6}$ \\
  0.5 & $70 \times 10^{-6}$ & $1.4 \times 10^{-6}$ \\ \hline \hline
\end{tabular}
\end{table}

The pattern of interference of contributions to the penguin amplitude of the
nonstrange and strange quarks is reversed for $B \to K^* (\eta,\eta')$ in
comparison with that for $B \to K (\eta,\eta')$ \cite{Lip87}.  Thus, one gets
contructive interference in $B \to K^* \eta$ and destructive interference in
$B \to K^* \eta'$.  With the mixing pattern adopted above, one has (again
neglecting small tree contributions to $B^+$ decays)
$$
A(B \to \eta K^*) = \frac{2}{\st}P_{PV} + \frac{1}{\st} S_{PV}~~,
$$
$$
A(B \to \eta' K^*) = - \frac{1}{\st}P_{PV} + \frac{4}{\st} S_{PV}.
$$
Here it is the $\eta K^*$ decay which is seen, with a branching ratio of about
$2 \times 10^{-5}$ \cite{Brau}.  If the singlet penguin contribution $S_{PV}$
were neglected, one would have $\Gamma(\eta' K^*) = (1/8) \Gamma(\eta K^*)$,
in the absence of phase space suppression, whose inclusion leads to the
prediction $\b(B \to \eta' K^*) = 2 \times 10^{-6}$.  Deviation from this
prediction would indicate evidence for the singlet penguin \cite{VPUP}.

The sign flip just mentioned arises from the behavior under charge conjugation
when one compares penguin amplitudes in which the spectator quark ends up in
the $\eta$ or $\eta'$ with those in which the $K^{(*)}$ contains the spectator.
Other evidence for a similar sign flip arises in the differing relative phases
of $I = 1/2$ and $I=3/2$ amplitudes in $D \to K^* \pi$ and $D \to K \rho$
\cite{JRcharm}, and in the observation \cite{Lip02} that $\b(D^+ \to K^{*+}
\bar K^0) = (3.1 \pm 1.4)\%$ shows an enhancement while $\b(D^+ \to K^+
\bar K^{*0}) = (0.42 \pm 0.05)\%$ does not, presumably because of constructive
vs.\ destructive interference of ``annihilation'' and ``tree'' amplitudes.

\subsection{$B_s$--$\bar B_s$ mixing and CKM elements}

Experimental \cite{TMoore} and theoretical \cite{Parodi} aspects of
$B_s$--$\bar B_s$ mixing may be summarized by noting that the present
lower limit $\Delta m_s \ge 14.9$ ps$^{-1}$ was not expected until recently
to be much below the actual value.  This was based on the estimate
$$
\frac{\Delta m_s}{\Delta m_d} = \xi^2 \frac{m_{B_s}}{m_{B_d}}~~,~~~
\xi \equiv \frac{f_{B_s} \sqrt{B_{B_s}}}{f_B \sqrt{B_B}}~~~,
$$
with lattice gauge theory estimates in the vicinity of $\xi \simeq 1.14 \pm
0.06$.  However \cite{Kron}, the error on $\xi$ may have been underestimated
in applying chiral perturbation theory when extrapolating to light masses
of the $u,d,s$ quarks; the new value is $\xi = 1.30 \pm 0.10$, which for
fixed Wolfenstein parameters $(\rho, \eta)$ changes $\Delta m_s$
by $+ 30\%$, allowing values up to 30 ps$^{-1}$.  For fixed $\Delta m_s$
it allows larger values of $|V_{td}|$, $|1 - \rho - i \gamma|$, and $\gamma$.

A quark model estimate \cite{JRFM} used the near-equality of hyperfine
splittings in the $D_s^*$--$D_s$ and $D^*$--$D$ systems to estimate
$f_{D_s}/f_D \simeq 1.25$, with the relation \cite{Grin} $f_{B_s}/f_B
\simeq f_{D_s}/f_D$ then yielding a similar ratio for $f_{B_s}/f_B$.  There
are good prospects for measuring $f_{D_s}/f_D$ at CLEO-c \cite{Pedlarc}.

\subsection{Angles of the unitarity triangle}

The angles of the unitarity triangle were called $\phi_1$, $\phi_2$, and
$\phi_3$ (opposite the sides corresponding to the products $V^*_{ib}V_{id}$
for the first, second, and third families) in 1987 \cite{HRS}.  These
``hiragana'' names, used by the Belle \cn, are expressed in ``katakana'' by
the BaBar \cn~as $\alpha \equiv \phi_2$, $\beta \equiv \phi_1$, and $\gamma
\equiv \phi_3$.

D. Marlow \cite{Marlow} reported a world average $\sin (2 \phi_1) = 0.78 \pm
0.08$ based on the CP asymmetries in $B \to J/\psi K_S$ and closely related
decays, driven mainly by the BaBar and Belle data.  These have subsequently
been updated, with BaBar \cite{Ba0702} now reporting $\sin (2 \phi_1) =
0.741 \pm 0.067 \pm 0.033$ and Belle \cite{Be0702} reporting $\sin (2 \phi_1) =
0.719 \pm 0.074 \pm 0.035$.  My average of these two values is $0.731
\pm 0.055$.

The best information on $\phi_2 = \alpha$ comes from $B^0 \to
\pi^+ \pi^-$, whose penguin amplitude (about 0.3 of the dominant tree)
complicates the analysis.  One can obtain the penguin from $B^+ \to
K^0 \pi^+$ (where it dominates) with the help of flavor SU(3) and an
estimate of symmetry-breaking.  Time-dependent asymmetries in this process
are proportional to $S_{\pi \pi} \sin(\Delta m t) - C_{\pi \pi} \cos(\Delta m
t)$.  The ``indirect'' $S_{\pi \pi}$ term is proportional to $\sin(2
\phi_{2{\rm eff}})$, where $\phi_{2{\rm eff}} \to \phi_2$ in the limit of a
vanishing penguin amplitude.  The ``direct'' $C_{\pi \pi}$ term is proportional
to the sine of a strong phase difference between penguin and tree amplitudes,
expected to be small in the generalized factorization approach \cite{Ben02}.
As pointed out by Morii \cite{Morii}, present data are beginning to constrain
$\phi_2$.  Datta \cite{Datta} (reporting on work in collaboration with D.\
London) has noted that analysis of several $B^0 \to K^{(*)0} \bar
K^{(*)0}$ modes may allow a clean determination of $\phi_2$.

The angle $\phi_3 = \gamma$ is harder to pin down in a model-independent way,
though there are ways to measure it to about $\pm 10^\circ$ using
various manifestations of tree-penguin interference \cite{VPUP,He,GRTP}.
These occur in such processes as $B \to \pi \pi$, $B_{(s)} \to K \pi$,
$B_s \to K \bar K$, and modes involving one light pseudoscalar and one
light vector meson.  For these (as well as for the modes $B^+ \to \pi^+ \eta$
and $B^+ \to \pi^+ \eta'$ which exhibit direct CP asymmetries) it will be
necessary to measure branching ratios with an accuracy of $\pm$ (1--2) $\times
10^{-6}$.

\subsection{SCET and factorization}

The ``generalized factorization'' approach we have mentioned has been
simplified considerably thanks to work by Stewart and collaborators
\cite{Stewart}, who developed a technique known as the soft collinear effective
theory (SCET).  They are able to prove factorization to all orders in
$B^0 \to D^- \pi^+$.  Corrections of order $1/m_c$ are responsible for the
fact that $\b(\bar D^0 \pi^+) \simeq 1.85 \b(D^- \pi^+)$.  They are
comfortable with the range of $\delta_I = {\rm Arg}(A_{3/2}/A_{1/2})$ reported
by CLEO \cite{PedlarD,Ahmed}.  An investigation of $B^0 \to \pi^+ \pi^-$ is in
progress.  At stake is whether the strong relative phase between tree and
penguin amplitudes is small (e.g., \cite{Ben02}) or large (e.g., \cite{KLS}).
At the moment experiment is no help in deciding this question, since BaBar
and Belle report very different values of the direct asymmetry parameter
$C_{\pi \pi}$, as shown in Table 4.

\begin{table}[h]
\caption{Time-dependent asymmetry parameters in $B^0 \to \pi^+ \pi^-$.}
\begin{tabular}{c c c} \hline \hline
Expt. & $S_{\pi \pi}$ & $C_{\pi \pi}$ \\ \hline
BaBar &    $0.02$    &     $-0.30$   \\
\cite{Bapp} & $\pm 0.34 \pm 0.05$ & $\pm 0.25 \pm 0.04$ \\ \hline
Belle &    $-1.21$    &     $-0.94$   \\
\cite{Bepp} & $^{+0.38+0.16}_{-0.27-0.13}$ & $^{+0.31}_{-0.25} \pm 0.09$ \\
\hline \hline
\end{tabular}
\end{table}

One point in favor of small $C_{\pi \pi}$
\cite{He} is the flavor-SU(3) relation $\Delta(B^0 \to \pi^+ \pi^-) =
- \Delta(B^0 \to K^+ \pi^-)$, where $\Delta(B \to PP) \equiv \Gamma(B \to
PP) - \Gamma(\bar B \to \bar P \bar P)$.  Since the direct CP asymmetry in
$B^0 \to K^+ \pi^-$ is small, one may also expect it to be small in
$B^0 \to \pi^+ \pi^-$. 

Applying the SCET, Leibovich \cite{Leib} reported on a study of $\Upsilon$
radiative decays near the end point, where one can justify the use of a
nonperturbative ``shape function'' near $z \equiv 2 E_\gamma/M_\Upsilon = 1$.
The effect of the color-octet admixture in the $\Upsilon$ wave function is
found to be surprisingly small.  Calculations are now in progress for the
color-singlet component of the wave function.

Colangelo \cite{Col} reported on a study of three-body
$B^0 \to D^{*-} D^{(*)0} K^+$ decays in which one can select the effects of
contributing $D_s^{(*)+}$ poles; factorization (like Niels Bohr's horseshoe)
works even when it isn't supposed to, as has been found elsewhere
\cite{JRFM,LRfact}.

\subsection{$B \to (s \gamma, s l^+ l^-, l^+ l^-)$ decays}

The experimental situation on radiative $b$ decays continues to improve.
Brau \cite{Brau} reported on a branching ratio $\b(B \to X_s \gamma) =
(3.22 \pm 0.40) \times 10^{-4}$, to be compared with the standard model
prediction \cite{Hiller,Gam} of $(3.54 \pm 0.49) \times 10^{-4}$.  A lot of the
theoretical uncertainty arises from uncertainty in $m_c/m_b$.  With
500 fb$^{-1}$ the experimental error is anticipated to be $\pm 1.8\%$,
making $\pm 3\%$ a useful theoretical goal.  BaBar has an upper limit
on $B \to \rho \gamma$ entailing $|V_{td}/V_{ts}| < 0.36$, approaching
the SM level of 0.2.

Both BaBar \cite{Brau} and Belle \cite{BellK} now see a $B^\pm \to K^\pm l^+
l^-$ signal, with BaBar newly reporting $\b = (0.84^{+0.30+0.10}_{-0.24-0.18})
\times 10^{-6}$.  Belle's inclusive branching ratios $\b(B \to X_s \mu^+
\mu^-) = (8.9^{+2.3+1.6}_{-2.1-1.7}) \times 10^{-6}$ and $\b(B \to X_s l^+
l^-) = (7.1 \pm 1.6 ^{+1.4}_{-1.2}) \times 10^{-6}$ are a bit above the
SM predictions \cite{Hiller} of $\b(B \to X_s e^+ e^-) = (6.9 \pm 1.0) \times
10^{-6}$ and $\b(B \to X_s \mu^+ \mu^-) = (4.2 \pm 0.7) \times 10^{-6}$.

In a discussion of radiative $b$ decays and related gateways to new physics,
Hiller \cite{Hiller} listed the ``Top 10 observables beyond $b \to s \gamma$.''
Briefly, these are:  (1) CP violation and (2) photon helicity in $b \to s
\gamma$, (3) $\sin(2 \phi_1)$ in $B \to \phi K$, (4) the dilepton mass spectrum
and (5) the forward-backward asymmetry in $b \to s l^+ l^-$, (6) the
question of where and whether this asymmetry vanishes as a function of dilepton
mass, (7) CP violation in this forward-backward asymmetry, (8) $B_s$--$\bar
B_s$ mixing and the effects of $Z$ penguins, (9) Higgs boson exchange in
$B \to l^+ l^-$, and (10) the neutron electric dipole moment. Geng \cite{Geng}
has suggested looking for new physics in T-violating observables in $\Lambda_b
\to \Lambda e^+ e^-$; P. Cooper pointed out in the discussion that one should
first look for the much more abundant process $\Sigma^+ \to p e^+ e^-$.
Huang \cite{Huang} notes that an interesting level for $\b(B^0 \to l^+ l^-)$
to display non-standard physics is in excess of $2 \times 10^{-8}$, which
sets an initial goal for high-statistics studies of this low-background
process.

\section{HEAVY QUARK PRODUCTION}
\label{sec:hqp}

The fragmentation of $c$ and $b$ quarks produced in $Z$ decays turns out
to be an important source of charmed and $b$ flavored baryons.  The DELPHI
\cn~\cite{Bern} has presented evidence for $c \to \Xi_c^0 \to \Xi^- \pi^+$
and $b \to \Xi_b \to \Xi^- l X$ with branching ratios $\b \sim 5$--$6 \times
10^{-4}$.

The color-octet part of the quarkonium wave function seems to be needed to
explain $J/\psi$ production at the Tevatron, but it is not so clearly required
in some other cases \cite{Brugnera}.  It is also claimed to be needed in
the description of decays to light hadrons (see, e.g., \cite{Maltoni}),
raising the question of how the value of $\alpha_s(m_b)$ extracted in older
analyses (e.g., \cite{KMRR}) would be affected.  In production models,
one of the hardest things to get right is the $J/\psi$ polarization,
reminiscent of the difficulties that Regge pole fits in the 1960s had in coping
with polarization data.

The production cross section for $b$ quarks at the Tevatron \cite{Cranshaw}
(and also in $ep$ \cite{Gerhards} and $\gamma \gamma$ reactions) exceeds the
predictions of non-leading-order (NLO) QCD, by as much as a factor of 2.5
in the case of $\bar p p$ collisions at 1.8 TeV.  Is this due to a subtlety
in the choice of QCD scale or an indication of new physics?  One proposed
scenario \cite{Berger} involves a light gluino and $b$ squark, whereby 
gluino pair production boosts the $b$ quark production cross section.
It seems difficult to rule out this scenario based on the $Q^2$ dependence
of $\alpha_s$ \cite{CLRas}.  Hadronic and $ep$ charm production, paradoxically,
seems less out-of-line \cite{Gerhards}.

Leading-quark effects in hadronic charm production have been studied by the
SELEX \cn~\cite{Iori}.  The presence of a specific quark or antiquark in the
beam governs the nature of the leading $D$ or $\bar D$, which will contain
that quark.  An interesting difference between $x_F$ distributions occurs
between $D^+$ and $D^{*+}$ produced by $\Sigma^-$ beams.

The HERA-b \cn~has measured the $b$ production cross section for $920$ GeV
protons on a fixed nuclear target: $\sigma(b \bar b) = 32^{+14+6}_{-12-7}$
nb/nucleon \cite{Mevius}.  At this energy, roughly 1/3 of prompt $J/\psi$
particles come from $\chi_c \to J/\psi \gamma$.

In a {\it tour de force} of track finding and scanning, the CHORUS \cn~has
measured $D^0$ and $\Lambda_c$ yields in charged-current neutrino
interactions at average beam energy 27 GeV \cite{Narita}:
$$
\frac{\sigma(D^0)}{\sigma_{cc}} = (1.99 \pm 0.13 \pm 0.17)\%~~,
$$
$$
\frac{\Sigma(\Lambda_c)}{\sigma_{cc}} = (1.39 \pm 0.18 \pm 0.27)\%~~.
$$
A cut $p_\mu < 30$ GeV was imposed in
order to ensure sufficient hadronic boost in the hybrid emulsion detector.

Heavy-ion hyperon and charm production has been studied at CERN
\cite{Safarik}.  There appears to be an enhancement of hyperon production
in nucleus-nucleus collisions, while an abrupt suppression of $J/\psi$
production occurs as the atomic number of the colliding particles is
increased, suggesting the onset of production of a quark-gluon
plasma in which quarkonium is dissociated.

\section{ELECTROWEAK SECTOR}
\label{sec:ew}

A persistent problem in fits to precision electroweak data has been the
forward-backward asymmetry in $e^+ e^- \to b \bar b$ \cite{Liebig}.  It is
given in terms of couplings of left-handed and right-handed fermions $i$
to the $Z$ by
$$
A^b_{FB} = \frac{3}{4}A_b A_e~~,~~A_i \equiv \frac{g^2_{Li}-g^2_{Ri}}
{g^2_{Li}+g^2_{Ri}}~~.
$$
In the SM one has $g_{Le} = -(1/2) + \sst$, $g_{Re} = \sst$, $g_{Lb} =
-(1/2) + (1/3) \sst$, $g_{Rb} = (1/3) \sst$.  The observed value of
$A^b_{FB}$ entails $\sst = 0.23218 \pm 0.00031$, to be compared with
$0.23149 \pm 0.00017$ in the overall fit.  This could be interpreted as a
discrepancy in $g_{Rb}$, with the data implying $0.095 \pm 0.008$ to be
compared with $\simeq 0.077$ in the SM.  However, $g_{Lb}$ seems to agree
roughly with its SM value of $\simeq -0.42$.  Other data, such as the
polarization asymmetry $A_{LR}$ measured at SLD \cite{deGroot}, suggest
a lower value of $\sst$, entailing a value of the Higgs boson mass $M_H$
lower than the experimental lower bound!  (See \cite{Chan} for a discussion.) 
Above the $Z$ \cite{Vorobiev}, there seem to be no anomalies in electroweak
couplings.  The $\tau$ lepton appears to behave as a standard-model
sequential fermion \cite{Sobie}.

If one also includes recent
data from the NuTeV \cn~\cite{NuTeV} on the neutral current cross section
$\sigma_{NC}(\nu N)$ for deep-inelastic scattering of neutrinos on nucleons,
which entail a {\it higher} value of $\sst$, the overall quality of the fit to
electroweak data is degraded, but the bound on $M_H$ is relaxed
\cite{JRAPV,Holzner}.  Higgs boson searches will tell us whether this is the
correct alternative.  At LEP \cite{Holzner},
only ALEPH claims a signal ($3 \sigma$, 115.6 GeV), diluted to $2.1 \sigma$
when data from DELPHI, L3, and OPAL are included.  The measurement of
four-fermion production at LEP II \cite{Boeriu} is instrumental in
understanding backgrounds.  Inclusion of NuTeV data in
electroweak fits relaxes the upper bound on $M_H$ to 212 GeV, assuming only
Higgs doublets acquire vacuum expectation values (VEVs).  With less than a 3\%
admixture of weak-SU(2) triplet VEVs, this bound is removed for all practical
purposes \cite{JRAPV}.

\section{BEYOND THE STANDARD MODEL}
\label{sec:bsm}

The reigning candidate for physics beyond the Standard Model is supersymmetry,
a beautiful concept which implies that there may be quantities
more fundamental than space and time.  Its implementation at the
electroweak scale is problematic for model-builders, perhaps because of
a lack of experimental clues.  I am tempted to ask, as A. Pais used to do
on other occasions, ``Where's the joke?''  Are we simply going to be
confronted with a whole new set of superpartner masses and mixings as
inexplicable as those of the quarks and leptons?  Or will supersymmetry help
us understand something about the pattern of quark and lepton families?

A very nice review of LEP searches for supersymmetry was
presented by B. Clerbaux \cite{Clerbaux}.  I am partial to one scheme based
on the grand unified group E$_{\rm 6}$ that is at least partly supersymmetric.
Start with the grand unified group SO(10), in which each quark and lepton
family -- including a right-handed neutrino -- is represented by a
16-dimensional spinor.  The simplest scalars in this scheme correspond to the
10-dimensional vector representation.  Superpartners of these states would
be 16-dimensional scalars (squarks and sleptons) and 10-dimensional
fermions (higgsinos).  By adding one more SO(10) singlet per family,
one can form families which belong to the 27-dimensional (fundamental)
representation of E$_{\rm 6}$.  The three SO(10) singlet fermions correspond to
sterile neutrinos.  The E$_{\rm 6}$ group fits nicely into superstring schemes,
and has been discussed recently by Bjorken, Pakvasa, and Tuan \cite{BPT}.

Present $e^+ e^-$ searches for supersymmetry (or for any particles coupling
with at least electroweak strength to the $\gamma$ and $Z$) typically
exclude most superpartners with masses less than the beam energy minus a
small model-dependent amount.  Unsurprisingly, lower limits at the Tevatron
tend to be higher but require a larger mass difference between the
next-to-lightest and lightest superpartner.

\section{PROSPECTS}
\label{sec:pros}

Future experiments on beauty, charm, and hyperons hold great promise.  BaBar
and Belle each are approaching integrated luminosities of 100 fb$^{-1}$ and
expect to have 500 fb$^{-1}$ by 2006.  The CDF and D0 detectors have new
capabilities for studying $b$ physics with high statistics and may have a
shot at the Higgs boson \cite{Cranshaw,Miller,Jesik,RMoore}.  The CLEO-c
project \cite{Pedlarc} is
on track to make a major impact on charm studies in the next few years.  In the
longer run, the forward special-purpose detectors BTeV \cite{Kas,Kwan} and
LHCb \cite{Aja} will be designed to take advantage of the relatively
large hadronic $b$ production cross section while suppressing backgrounds.
The general-purpose ATLAS and CMS detectors at the LHC also will have many
capabilities related to topics discussed at this Conference
\cite{Nairz,Ohl,Ran,Seg,Smiz,Speer}.

Examples of questions I would like to see answered are the following.

(1) What value of $\phi_2 = \alpha$ is implied by the indirect CP asymmetry in
$B^0 \to \pi^+ \pi^-$?  The BaBar-Belle discrepancy should be resolved.
The branching ratio $\b(B^0 \to \pi^+ \pi^-)$ is itself of interest.  It can
be combined with information on $B \to \pi l \nu$, using factorization, to
estimate the effects of tree-penguin interference \cite{LRpipi}.

(2) What is $\Delta m_s$?  This will be an early Tevatron result
if all goes well.

(3) Can we obtain a value of $\phi_3 = \gamma$ by studying such decays
as $B^0 \to \pi^+ \pi^-$, $B_s \to K^+ K^-$, and $(B^0~{\rm or}~B_s) \to
K^\pm \pi^\mp$, or will symmetry-breaking and rescattering effects prove
uncontrollable?  Measurements of branching ratios at the level of
$10^{-7}$ will help settle these questions.

(4) Can we exploit $B_s \to J/\psi \phi$ by employing
transversity analyses to separate out CP-even and CP-odd final states?
The CP eigenstates $J/\psi \eta$ or $J/\psi \eta'$ may also be useful.  The
Standard Model predicts the CP asymmetries in these modes to be small but there
could always be surprises.

(5) If we see non-standard physics (e.g., in
$B \to \phi K_S$), will we able to identify it?  Hiller's ``top 10'' list
\cite{Hiller} is an interesting starting point; what then?

(6) If we see a Higgs boson, will we be able to tell whether it is the
Standard Model or supersymmetric variety?

(7) Are we sufficiently attuned to the wide variety of new things we could see
(supersymmetry, extra dimensions, exotic quarks and leptons, $\ldots$)
at Tevatron Run II?

(8) What will the LHC reveal (and when)?

\section*{ACKNOWLEDGMENTS}

The smooth arrangements made by our hosts have made us all feel exceptionally
welcome at UBC.  I would like to thank the Conference Assistants:
Travis Beale, Patrick Brukiewich, Janet Johnson, Mark Laidlaw, and Douglas
Thiessen, for their efforts, and the Organizing Committee:  Janis McKenna, Tom
Mattison, John Ng, Marco Bozzo, Calvin Kalman, Zoltan Ligeti, Miguel-Antonio
Sanchis-Lozano, and Paul A. Singer, for putting together a varied and
informative program.  We in Chicago look forward to welcoming you to our city
for the Sixth Conference on Beauty, Charm, and Hyperons in 2004.

I am grateful to the Theory Groups of Fermilab and Argonne National Laboratory
for hospitality during the preparation of the written version of this talk, and
to Matthias Neubert for useful discussions.  This work was supported in part by
the United States Department of Energy through Grant No. DE FG02 90ER40560.

\end{document}